\documentclass[preprint,12pt]{elsarticle}

\usepackage{amssymb}
\usepackage{color}
\usepackage{graphicx}
\usepackage{multirow}
\usepackage{tabularx}
\usepackage{booktabs}
\usepackage{hyperref}
\usepackage{subfigure}
\usepackage{slashbox}
\usepackage[ruled,lined,algonl]{algorithm2e}
\usepackage{amsmath}
\usepackage[figuresright]{rotating}
\usepackage{setspace}
\journal{Nuclear Physics B}

\begin{document}

\begin{frontmatter}



\title{Using Text Mining To Analyze Real Estate Classifieds}

\author[a,b]{Sherief Abdallah\corref{c}}
\address[a]{British University in Dubai, United Arab Emirates}
\address[b]{University of Edinburgh, United Kingdom}
\ead{shario@ieee.org}
\cortext[c]{Corresponding author}

\begin{abstract}
 Many brokers have adapted their operation to exploit the potential of the web. 
Despite the importance of the real estate classifieds, there has been little work in analyzing such data. 
In this paper we propose a two-stage regression model that exploits the textual data in real estate classifieds. We show how our model can be used to predict the price of a real estate classified. We also show how our model can be used to highlight keywords that affect the price positively or negatively. To assess our contributions, we analyze four real world data sets, which we gathered from three different property websites. The analysis shows that our model (which exploits textual features) achieves significantly lower root mean squared error across the different data sets and against variety of regression models.
\end{abstract}

\end{frontmatter}

\section{Introduction}

Real estate classifieds constitute an integral part of the real estate market. 
A real estate classified provides a concise description of a real estate unit that is available either for rent or sale. Traditionally, such classifieds were publicized using printed materials such as newspapers or dedicated classifieds periodicals. 
The Internet revolutionized the classifieds business, as with other advertisement sectors. The large reduction in price, coupled by efficient and quick search service, removed the restriction on classifieds size (from a maximum of few tens of words to practically unlimited textual description) and allowed the inclusion of multi-media images and videos. Such revolution can be verified through the valuation of Zillow, one of the biggest websites that specializes in posting real-estate classifieds, at 50 billion USD.\footnote{"Hedge fund sees Zillow becoming a \$50B company", \url{http://www.inman.com/2014/11/19/zillows-biggest-investor-expects-portal-to-reach-50b-valuation/} accessed pm 1st of September 2015.} 
Although at some point it was feared that such a new trend in advertising properties may threaten the profitability of traditional brokerage companies, many brokers have adapted their operation to exploit what the web has to offer; and they integrated the web technology into their process \cite{crowston99ijem}. 

Nowadays, many large brokerage companies developed their own websites to list their properties, in addition to listing their properties on 3rd party web portals. In United Arab Emirates, where the real estate market constitutes 12.5\% of the GDP,\footnote{"Dubai’s GDP climbs 4.4\%", Khaleej Times, 12th June 2013.} several major website portals targeted real estate classifieds. Examples include Gulf News Ads (www.gnads4u.com), Dubizzle (www.dubbizle.com), Bayut (www.bayut.com), and Property Finder (www.propertyfinder.ae). Furthermore, major real estate brokering companies list classifieds on their own websites, such as  such as Better Homes (www.bhomes.com) and Hamptons (www.hamptons.ae).

Despite the importance of the real estate classifieds, there has been little work in analyzing such data. In particular, most of the previous work that applied data mining to real estate data focused on structured attributes such as number of bedrooms, area, location, etc. \cite{guan08jrer} (a broader view of related work is given in Section \ref{sec-related}). For example, Zillow website provides Zestimate service which attempts to automatically valuate a real estate unit. The service relies on history of previously sold houses, using only structured features \cite{humphries2012automatically}.

However, for classifieds, the unstructured and ungrammatical\footnote{Ungrammatical means the text does not strictly conform to grammatical rules.} textual attributes (typically the classified title and description) are important components of a classified that should not be ignored as they may encapsulate ad hoc features (such as "remodeled house", "near X hotel"). While many websites try to capture ad hoc features in a form of a checklist that ad posters can mark, it is hard to account for all possible features in such manner, particularly rare features. Also some ad hoc features may appear suddenly over time (e.g. due to the opening of a new park or a hotel).

In this work we investigate the use of text mining, along with other data mining techniques, to analyze real estate classifieds.  Our aim is to explore two important research questions: 
\begin{itemize}
\item Can the use of text mining improve the accuracy of predicting the price of a real estate classified?  
\item Can we identify which keywords affect the price of a real estate unit either positively or negatively?
\end{itemize} 

Answering these questions will be valuable for real estate agents, and even home owners who directly post classifieds. Predicting a more accurate price, for a real estate unit, that takes into account the textual unstructured data (not just the structured data) will prevent the stakeholder from overestimating or underestimating the price. Furthermore, by identifying the important keywords, the stakeholder can refine the unit description and title to better reflect the price being asked. 

We propose a two-stage regression model to solve both of these questions. The first stage of our model uses only structured features of a classified (such as the number of bedrooms and the location) to make an initial prediction of the real estate unit. The second stage uses the textual features of the same classified to refine the initial prediction.
We conducted the analysis on four data sets of real estate classifieds. The datasets were gathered from three different websites that post real estate classifieds in United Arab Emirates. We show that our proposed approach improves the prediction of a property price or rental. We also illustrate how text mining combined with a linear regression model can be used to identify keywords that affect the price negatively or positively. Table \ref{tab-ad} shows an example classified that was \emph{automatically highlighted} using our proposed system.

\begin{table}[h!]
\small
\center
\fbox{
\parbox[l]{\textwidth}{
\texttt{
 2 \textcolor[rgb]{0.0,0.0,0.0}{BR}+\textcolor[rgb]{0.0,0.0,0.5450980392156862}{maid} \textcolor[rgb]{0.0,0.0,0.0}{with} \textcolor[rgb]{0.0,0.0,0.5333333333333333}{full} \textcolor[rgb]{0.0,0.0,0.0}{sea} \textcolor[rgb]{0.0,0.0,0.1803921568627451}{views} \textcolor[rgb]{0.0,0.0,0.0}{is} \textcolor[rgb]{0.0,0.0,0.06274509803921569}{available} \textcolor[rgb]{0.0,0.0,0.0}{for} \textcolor[rgb]{0.0,0.0,0.0}{rent} \textcolor[rgb]{0.0,0.0,0.0}{in} \textcolor[rgb]{0.0,0.0,0.0}{Al} \textcolor[rgb]{0.0,0.0,0.0}{Hasser}, \textcolor[rgb]{0.0,0.0,0.0}{Shoreline} \textcolor[rgb]{0.0,0.0,0.0}{Apartment}, \textcolor[rgb]{0.0,0.0,0.6235294117647059}{Palm} \textcolor[rgb]{1.0,0.0,0.0}{Jumeirah}. \textcolor[rgb]{0.0,0.0,0.13725490196078433}{Vacant} \textcolor[rgb]{0.0,0.0,0.0}{and} \textcolor[rgb]{0.0,0.0,0.0196078431372549}{ready} \textcolor[rgb]{0.0,0.0,0.0}{to} \textcolor[rgb]{0.12941176470588237,0.0,0.0}{move} \textcolor[rgb]{0.0,0.0,0.0}{in}.
\textcolor[rgb]{0.0,0.0,0.0}{The} 20 \textcolor[rgb]{0.0,0.0,0.0}{Shoreline} \textcolor[rgb]{0.0,0.0,0.0}{Apartment} \textcolor[rgb]{0.37254901960784315,0.0,0.0}{buildings} \textcolor[rgb]{0.0,0.0,0.0}{that} \textcolor[rgb]{0.0,0.0,0.0}{line} \textcolor[rgb]{0.0,0.0,0.0}{the} \textcolor[rgb]{0.0,0.0,0.0}{east} \textcolor[rgb]{0.0,0.0,0.0}{side} \textcolor[rgb]{0.0,0.0,0.0}{of} \textcolor[rgb]{0.0,0.0,0.0}{The} \textcolor[rgb]{0.0,0.0,0.0}{Trunk} \textcolor[rgb]{0.0,0.0,0.0}{feature} \textcolor[rgb]{0.0,0.0,0.0}{some} \textcolor[rgb]{0.0,0.0,0.0}{of} \textcolor[rgb]{0.0,0.0,0.0}{the} \textcolor[rgb]{0.0,0.0,0.0}{Middle} \textcolor[rgb]{0.0,0.0,0.0}{Easts} \textcolor[rgb]{0.0,0.0,0.0}{most} \textcolor[rgb]{0.0,0.0,0.0}{desirable} \textcolor[rgb]{0.6274509803921569,0.0,0.0}{apartments}. \textcolor[rgb]{0.0,0.0,0.0}{Five} \textcolor[rgb]{0.0,0.0,0.1411764705882353}{exclusive} \textcolor[rgb]{0.0,0.0,0.42745098039215684}{beachfront} \textcolor[rgb]{0.0,0.0,0.0}{clubhouses} \textcolor[rgb]{0.0,0.0,0.0}{cater} \textcolor[rgb]{0.0,0.0,0.0}{to} \textcolor[rgb]{0.0,0.0,0.0}{residents}, \textcolor[rgb]{0.0,0.0,0.0}{providing} \textcolor[rgb]{0.0,0.0,0.18823529411764706}{world} \textcolor[rgb]{0.0,0.0,0.0}{class} \textcolor[rgb]{0.0,0.0,0.0}{fitness} \textcolor[rgb]{0.0,0.0,0.0}{centres}, \textcolor[rgb]{0.0,0.0,0.0}{retail} \textcolor[rgb]{0.0,0.0,0.0}{outlets}, \textcolor[rgb]{0.0,0.0,0.0}{al} \textcolor[rgb]{0.0,0.0,0.0}{fresco} \textcolor[rgb]{0.0,0.0,0.615686274509804}{dining}, \textcolor[rgb]{0.0,0.0,0.0}{swimming} \textcolor[rgb]{0.23529411764705882,0.0,0.0}{pools} \textcolor[rgb]{0.0,0.0,0.0}{and} \textcolor[rgb]{0.0,0.0,0.0}{direct} \textcolor[rgb]{0.0,0.0,0.5372549019607843}{access} \textcolor[rgb]{0.0,0.0,0.0}{to} \textcolor[rgb]{0.0,0.0,0.0}{the} \textcolor[rgb]{0.0,0.0,0.3215686274509804}{islands} \textcolor[rgb]{0.0,0.0,0.0}{white} \textcolor[rgb]{0.0,0.0,0.0}{sand} \textcolor[rgb]{0.0,0.0,0.42745098039215684}{beaches}.
\textcolor[rgb]{0.0,0.0,0.011764705882352941}{Facilities}: 5 \textcolor[rgb]{0.0,0.0,0.0}{Health} \textcolor[rgb]{0.0,0.0,0.0}{Clubs} ( 1 \textcolor[rgb]{0.0,0.0,0.0}{for} 4 \textcolor[rgb]{0.6274509803921569,0.0,0.0}{apartments} \textcolor[rgb]{0.0,0.0,0.0}{blocks}), \textcolor[rgb]{0.0,0.0,0.0}{Large} \textcolor[rgb]{0.0,0.0,0.0}{Swimming} \textcolor[rgb]{0.23529411764705882,0.0,0.0}{Pool}, \textcolor[rgb]{0.03137254901960784,0.0,0.0}{Modern} \textcolor[rgb]{0.0,0.0,0.0}{Gym}, \textcolor[rgb]{0.0,0.0,0.0}{Children}'\textcolor[rgb]{0.0,0.0,0.0}{s} \textcolor[rgb]{0.0,0.0,0.0}{playground}, \textcolor[rgb]{0.0,0.0,0.0}{and} \textcolor[rgb]{0.03529411764705882,0.0,0.0}{Restaurants}
}
}
}
  \caption{Sample classified with important words highlighted. Words in red affect the price of the classified negatively (e.g. \textcolor[rgb]{1.0,0.0,0.0}{Jumeirah}), while words in blue affect the price positively (e.g. \textcolor[rgb]{0.0,0.0,0.6235294117647059}{Palm}).}
\label{tab-ad}
\end{table}

To summarize, the contributions of this paper are:

\begin{itemize}
\item proposing a 2-stage regression model that exploits the textual features of a real estate classified to improve the price prediction of the real estate unit.
\item showing how our proposed model can be used to automatically highlight keywords that affect the real estate price positively or negatively
\item collecting 4 datasets of real estate classifieds and using the 4 datasets to evaluate our proposed model.
\end{itemize}

The rest of the paper is organized as follows. The following section describes how the data was collected and prepared. We then explain our proposed data mining process for predicting the price of real estate units using text mining and linear regression model. This is followed by the evaluation and analysis of our proposed approach using the collected data. We then discuss the related work and conclude our paper.

\section{Data Preparation}

We extracted our data from three major websites that post on-line residential real estate classifieds in the United Arab Emirates. The data was collected in the period from 17th of September to 6th of October, using our own web crawler.\footnote{While more sophisticated crawlers do exist,\cite{pera13is}, for our purposes we preferred a simple and efficient crawler.} The collected data contained both apartments (flats) and villas (houses) that are offered for either sale or rent. A total of 20,600 records were extracted. Table \ref{tab-features} illustrates the extracted features.

  \begin{table}[htbp]
\centering
        \begin{center}
        \begin{tabularx}{\textwidth}{|c|c|X|}
\hline
Name & Type & Description \\
\hline
Title	& Text	& Title of the classified. \\
\hline
Description	& Text	& Description of the property.\\
\hline
Beds & Integer & Number of bedrooms in a property \\
\hline 
Baths & Integer & Number of bedrooms in a property  \\
\hline 
Size & Integer & Built area of the property (in square feet) \\ 
\hline
Location & Nominal	& Each nominal value represents a neighborhood.\\
\hline
Price & Integer	& The renting or selling price of the property.\\
\hline
\end{tabularx}
\end{center}
\caption{\small The features of the data set.}
\label{tab-features}
\end{table}

The extracted data was then cleaned as follows. Duplicate classifieds (identical title and description) were removed to endup with 19644 records. Then outliers were removed as follows. First, the average rent/price per bedroom was calculated by deviding \textbf{Price} over \textbf{Beds  + 1} (the added one was needed in case of a studio apartment which has value \textbf{Beds} = 0). Records with unreasonably low/high price were removed according to the following conditions. For the rental properties we only accepted the properties that satisfy the condition $100000 \ge AverageRent \ge 10000$.\footnote{The rental price in Dubai is annual, and the currency is Arab Emirates Dirham (AED) (The Dirham is pegged to the US dollar at a rate of 3.67 AED per US dollar).} . The properties that are up for sale needed to satisfy the condition: $1000000 \ge AveragePrice \ge 100000$.
This resulted in further reduction in the number of records to 16008 divided as follows: 
\begin{itemize}
\item Apartments for rent: 8192 records
\item Houses for rent: 2105 records
\item Apartments for sale: 4599 records
\item Houses for sale: 1112 records.
\end{itemize}
Finally, all textual features were converted to lower-case.

\section{2-Stage Regression Model with Text-Mining}
\begin{sidewaysfigure}[htbp] 
    \centering
		\includegraphics[width=8in]{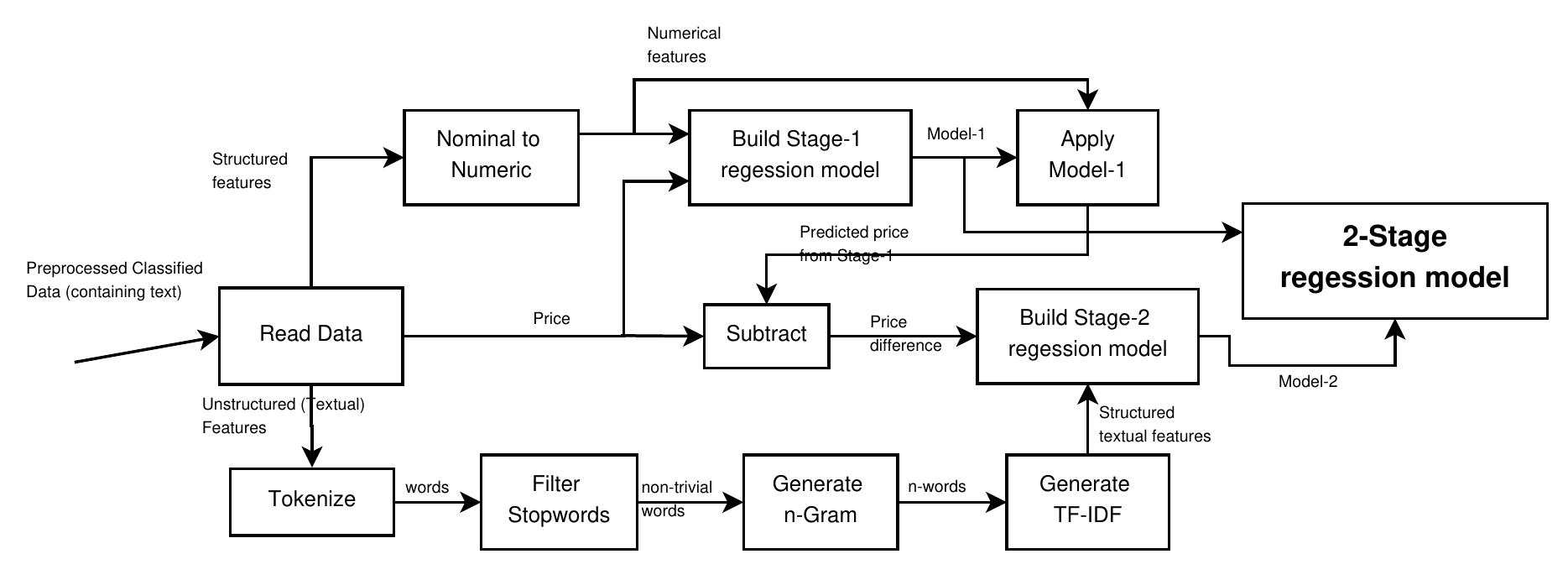}
	   \caption{\small The data mining process we used to build 2-stage regression model for real-estate classifieds.}
      \label{fig-process}
    \end{sidewaysfigure}

Figure \ref{fig-process} illustrates the process we propose to analyze real estate classifieds data. The main idea is to use a 2-stage regression model to predict the price/rent of a real estate classifieds. The first regression model (Stage-1 regression model in the figure) attempts to predict the price of a real estate classified using \emph{only structured features} (such as the number of bedrooms  and the location). The second regression model attempts to predict the remaining difference in price between the actual price and the predicted price from Stage-1 regression model. 

The decomposition of the regression of the price in two (sequential) stages achieves the following benefits:

\begin{description}
\item[Allowing different regression model for each stage]. This is important because while linear regression is convenient to use for the second stage to highlight which keywords are(not) important, linear regression is not necessarily the best regression model for the first stage.  
\item[Intuitive semantics]. With such decomposition, the semantics of the linear regression weights for the second stage are clear: explain why certain units have lower than (lager than) expected price, as predicted by the first regression model (which relies primarily on structured features), using keywords.
\item[Easier integration with state-of-the-art]. Since the current state-of-the-art techniques in automatic valuation of real estate units rely on structured features, our two stage approach allows easier integration with such techniques.
\end{description}

The process was implemented using RapidMiner.\footnote{RapidMiner is an open-source system for data mining that allows building a rich data-flow process of data-mining operators (http://rapid-i.com).} The following sub-sections describe the different components of our proposed approach in further detail. 

\subsection{Stage 1: Predicting Price Using Structured Features}
The first stage in our model uses structured features to predict the price of a real estate unit. The first step is converting the location attribute, which is nominal, to numerical attributes. This is done by creating a set of binary attributes, one binary attribute for each nominal value of the location.
 For example, "Dubai Marina" is a possible value of the location attribute. After this step we have a binary feature "loc\_dubai\_marina", which equals 1 if and only if the location attribute equals "Dubai Marina", otherwise it equals 0. 

The final step of Stage 1 is building a regression model for predicting the price using the structured (now numerical) features. We have considered 3 regression models in our evaluation: linear regression (LR), artificial neural networks (NN) \cite{hansen1990neural}, and support vector machine regression (SVMR) \cite{shevade2000improvements}. We describe LR in the following section because it is important to understand how we highlight important keywords. However, for the other two regression models, we refer the interested reader to the corresponding citation.

\subsection{Price Decomposition}
Before building the model for Stage 2, the predicted price in Stage 1 is calculated using the fitted model of Stage 1. Then the difference between the actual price and the price predicted by Stage 1 is calculated. This difference in price becomes the target attribute for Stage 2. In other words, Stage 2 attempts to predict the \emph{difference} between the actual price and the price predicted using only structured features.

\subsection{Stage 2: Using Text Mining to Improve Predicted Price}
The second stage converts the textual attributes to numerical features. As we mentioned earlier, The purpose of applying text-mining is to discover the effect of the hidden information in “Title” and “Description” features that might enhance the accuracy of predicting a property's price.
First, the text is tokenized by splitting the text into sequence of tokens, words. Tokens that are less than 4-characters length are removed. Then, stop-word tokens are removed using a list of English stop-words. A stop-word is a word that has little distinguishing power, such as "the", "is", and "at". This is followed by the generation of n-Grams terms. A term n-Grams is a series of successive $n$ tokens. For example, the set of 1-Gram terms is the set the original tokens. The set of 2-Gram terms contains sequences of two tokens and so forth. To avoid terms that too rare (possibly outliers) or too common (does not help in distinguishing classifieds), all terms that appear in less than $T_{min}$ percent or more than $T_{max}$ percent of all classifieds are removed.
The final step is computing the Term Frequency-Inverse Document Frequency (TF-IDF) for each term (in each classified). The TF-IDF counts how many times a particular term (n-Gram) $i$ appears in the text of classified $j$, which is then inversely weighted by how common the term is across different classifieds:

\[
\textrm{TF-IDF}( i, j ) = TF( i, j) * IDF(i)
\]  
where $TF(i,j) = \frac{F(i,j)}{|F|}$, $F(i,j)$ is the number of occurrences of Term $i$ in Classified $j$, $|F|$ is the length of the $F$ vector (F(i,j) $\forall i$), $IDF(i) = \log \frac{N}{N_i}$, $N$ is the number of documents (classifieds), and $N_i$  is the number of documents that contain Term $i$.

By the end of this step, the text of each classified is converted to a numerical vector representation. Each feature in the vector corresponds to a term, where the value is the TF-IDF of the term. The features are then filtered to reduce their numbers. We removed features that are highly correlated with one another (keeping only one feature from each set of correlated features). Two features are considered correlated if the correlation between them exceeds 0.99.
The final step of Stage 2 is building the regression model. Here we only use linear regression model because it assigns a weight for each term. LR follows the simple model:

\[
PredictedPrice( j ) = \sum_{i} w_i. a_i
\]

where $w_i$ is the weight corresponding to feature $i$ and $a_i$ is the value of feature $i$. The learning algorithm finds the best values for the weights based on the the dataset that minimizes (squared) error (between the actual price and the predicted price). The weight intuitively  reflects the feature's effect on the price. For example, when a feature has a positive weight value, it means that the feature works toward increasing the price. Similarly, having a feature with negative weight has the effect of decreasing the price.  We show in the following section how this intuition helps in identifying important keywords.

\subsection{Applying the Model}

\begin{sidewaysfigure}[htbp] 
    \centering
		\includegraphics[width=8in]{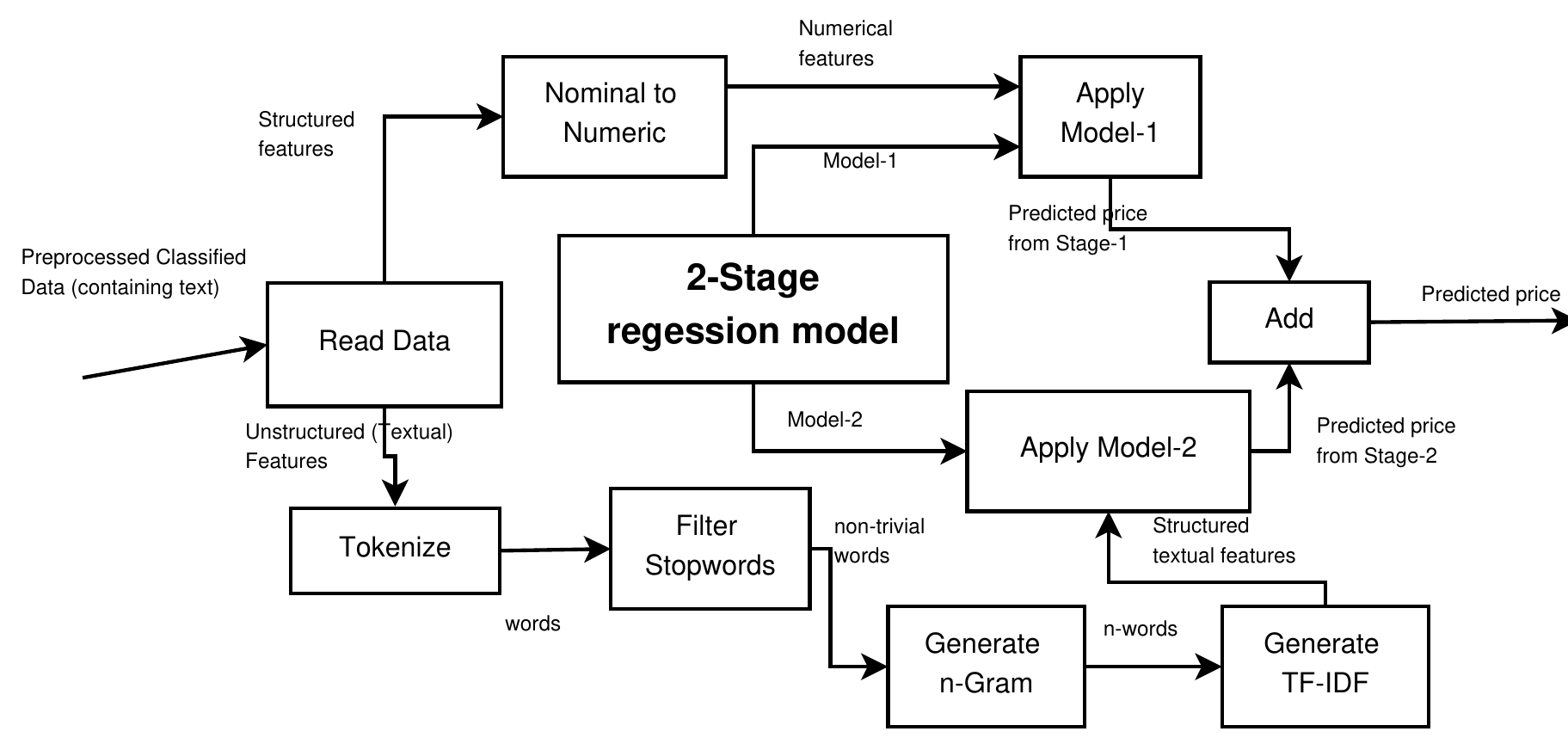}
	   \caption{\small The data mining process to apply the 2-stage regression model for real-estate classifieds and generating the predicted price.}
      \label{fig-apply}
    \end{sidewaysfigure}

Figure \ref{fig-apply} illustrates how the generated 2-stage regression model can be used to predict the price of a real estate classified. Each stage predicts a component of the price, and the two components are summed to generate the final prediction.

As we also showed in Table \ref{tab-ad}, the weights of the LR model of Stage 2 can be used to highlight important keywords.

\section{Analysis}

\begin{sidewaysfigure}[htbp] 
    \centering
		\includegraphics[width=8in]{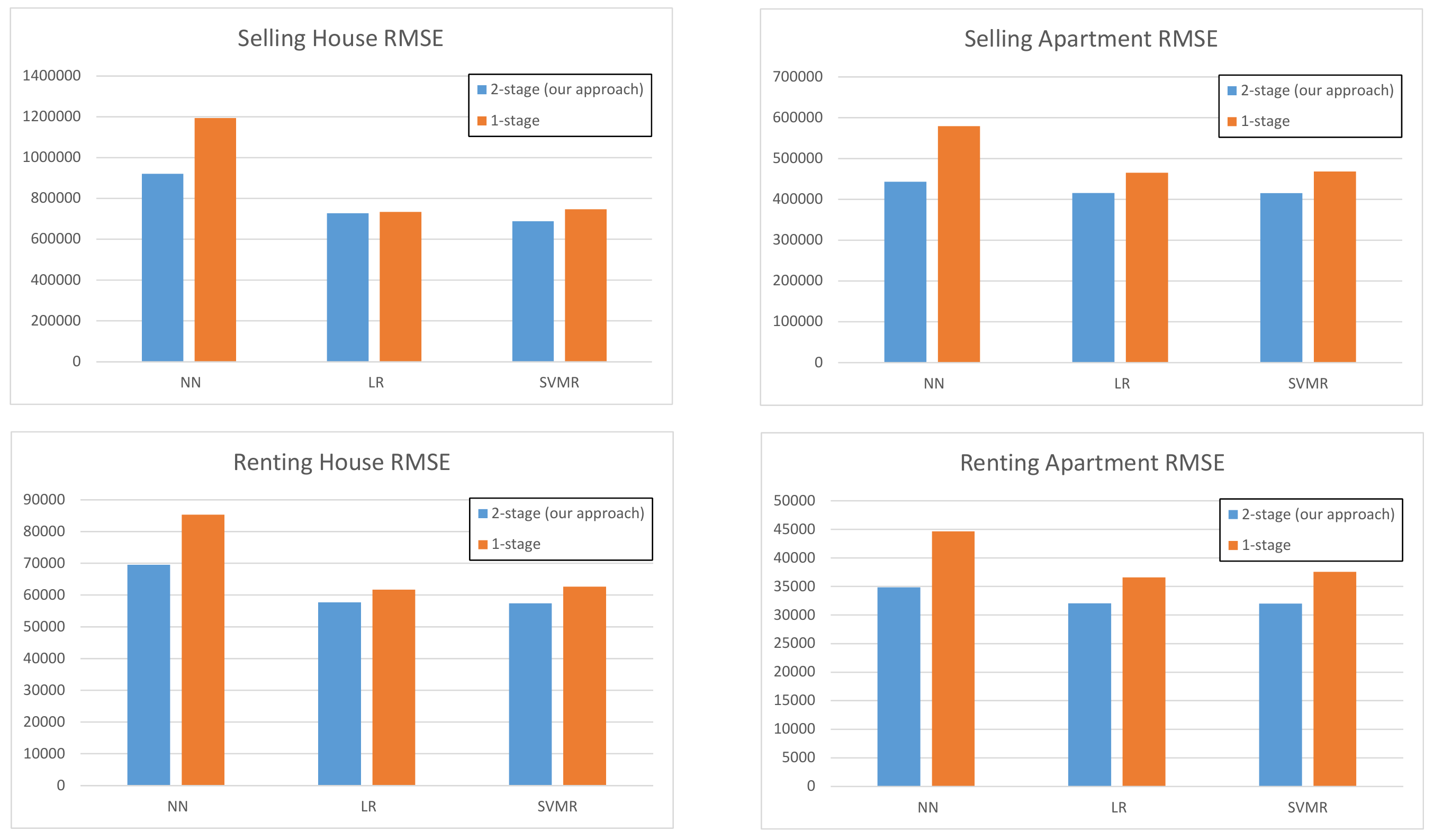}
	   \caption{\small Summary of results for the four data sets and the three regression models of Stage 1: selling house (SH), selling apartment (SA), renting house (RH), renting apartment (RA), linear regression (LR), artificial neural network (NN), and regression using support vector machine (SVMR).}
      \label{fig-results}
    \end{sidewaysfigure}

To evaluate our approach, we computed two performance metrics: the Root Mean Squared Error (RMSE) and the correlation between the predicted price and the actual price. These two measures were computed for each of the four datasets, with and without using the textual features. In other words, we compared our 2-stage regression model (which incorporated text mining) to the traditional (1-stage) regression model (which relied only on structured features). The comparison is done using three different regression models for the first stage while linear regression was used for the second stage. Figure \ref{fig-results} summarizes our results for the 4 datasets and the 3 regression models of Stage 1. In the remainder of this section we discuss in more detail the results for each regression model of Stage 1.

Table \ref{tab-res-lr-rmse} and Table \ref{tab-res-lr-corr} summarize the results when linear regression (LR) was used for the first stage. The tables   show a clear and statistically significant improvements, in all four datasets,  when using our 2-stage approach. The RMSE is reduced and the correlation increased. It is worth noting that the RMSE is a higher and the correlation lesser for houses when compared to apartments. This is due to the larger variety of houses even within the same neighborhood, in addition to the smaller number of  house classifieds (compared to apartments).

\begin{table}[htbp]
  \centering
  \caption{Root Mean Squared Error (RMSE) when linear regression was used in the Stage 1 of our 2-stage regression model.}
\begin{tabular}{|c|c|c|}
    \hline
    \textbf{Dataset} & \textbf{RMSE} w/o text-mining & \textbf{RMSE} with text mining \\
    \hline
    Renting apartment & 36737 +/- 1046 & 32047.045 +/- 1352.944 \\
    Renting house  & 61752 +/- 2814 & 57687.191 +/- 3176.731 \\
    Selling apartment & 465525 +/- 8399 & 415685.369 +/- 18292.775 \\
    Selling house & 735377 +/- 29075 & 727541.031 +/- 78869.986 \\
    \hline
    \end{tabular}%
  \label{tab-res-lr-rmse}%
\end{table}%

\begin{table}[htbp]
  \centering
  \caption{The correlation (between predicted and actual price) when linear regression was used in the first stage of our 2-stage regression model.}
\begin{tabular}{|c|c|c|}
    \hline
    \textbf{Dataset} & \textbf{Corr.} w/o text-mining & \textbf{Corr.} with text mining \\
    \hline
    Renting apartment & 0.838 +/- 0.012 & 0.881 +/- 0.008 \\
    Renting house  & 0.845 +/- 0.021 & 0.871 +/- 0.017 \\
    Selling apartment & 0.856 +/- 0.013 & 0.889 +/- 0.011 \\
    Selling house & 0.809 +/- 0.038 & 0.817 +/- 0.051 \\
    \hline
    \end{tabular}%
  \label{tab-res-lr-corr}%
\end{table}%

Table \ref{tab-res-nn-rmse} and Table \ref{tab-res-nn-corr} show similar results when we used artificial neural network (ANN) for the Stage 1. The tables   show  again a clear and statistically significant improvements when using our 2-stage approach against the traditional 1-stage approach (that also uses artificial neural network). The RMSE is reduced and the correlation increased in all four datasets. It is worth noting that across all datasets the performance of artificial neural network is worse than linear regression, but the use of text mining in Stage 2 still improved performance.

\begin{table}[htbp]
  \centering
  \caption{Root Mean Squared Error (RMSE) when artificial neural network regression was used in the first stage of our 2-stage regression model.}
\begin{tabular}{|c|c|c|}
    \hline
    \textbf{Dataset} & \textbf{RMSE} w/o text-mining & \textbf{RMSE} with text mining \\
    \hline
    Renting apartment & 44648.549 +/- 6795.781 & 34843.054 +/- 5676.464 \\
    Renting house  & 85353.454 +/- 26446.908 & 69535.376 +/- 6847.498 \\
    Selling apartment & 579014.731 +/- 120081.461 & 443262.976 +/- 24856.718 \\
    Selling house & 1193741.188 +/- 568449.993 & 920461.846 +/- 144612.999 \\
    \hline
    \end{tabular}%
  \label{tab-res-nn-rmse}%
\end{table}%

\begin{table}[htbp]
  \centering
  \caption{The correlation (between predicted and actual price) when artificial neural network regression was used in the first stage of our 2-stage regression model.}
\begin{tabular}{|c|c|c|}
    \hline
    \textbf{Dataset} & \textbf{Corr.} w/o text-mining & \textbf{Corr.} with text mining \\
    \hline
    Renting apartment & 0.811 +/- 0.034 &  0.871 +/- 0.027 \\
    Renting house  & 0.714 +/- 0.244 & 0.816 +/- 0.040 \\
    Selling apartment & 0.835 +/- 0.024  & 0.878 +/- 0.016 \\
    Selling house & 0.490 +/- 0.339 & 0.684 +/- 0.110  \\
    \hline
    \end{tabular}%
  \label{tab-res-nn-corr}%
\end{table}%

Table \ref{tab-res-svm-rmse} and Table \ref{tab-res-svm-corr} show similar results when we used support vector machine regression (SVMR) for the first stage of our approach. Similar to the previous regression models we evaluated, the tables  show a clear and statistically significant improvements when using our 2-stage approach against the traditional 1-stage approach (that also uses SVMR). Interestingly, using the traditional 1-stage approach, SVMR is slightly worse than LR. However, using our approach and exploiting the textual features of ads, SVMR results in the lowest RMSE and the highest correlation in all four datasets with a clear advantage over linear regression for the Selling House dataset.

\begin{table}[htbp]
  \centering
  \caption{Root Mean Squared Error (RMSE) when support vector machine regression was used in the first stage of our 2-stage regression model.}
\begin{tabular}{|c|c|c|}
    \hline
    \textbf{Dataset} & \textbf{RMSE} w/o text-mining & \textbf{RMSE} with text mining \\
    \hline
    Renting apartment & 37550.226 +/- 1814.974  & 32003.126 +/- 1676.801 \\
    Renting house  & 62682.050 +/- 4487.318 & 57377.165 +/- 3531.160 \\
    Selling apartment & 467869.878 +/- 13612.336 & 415388.299 +/- 18916.617 \\
    Selling house & 746897.386 +/- 46691.263 & 688058.699 +/- 66977.813 \\
    \hline
    \end{tabular}%
  \label{tab-res-svm-rmse}%
\end{table}%

\begin{table}[htbp]
  \centering
  \caption{The correlation (between predicted and actual price) when support vector machine regression was used in the first stage of our 2-stage regression model.}
\begin{tabular}{|c|c|c|}
    \hline
    \textbf{Dataset} & \textbf{Corr.} w/o text-mining & \textbf{Corr.} with text mining \\
    \hline
    Renting apartment &0.833 +/- 0.014 &  0.880 +/- 0.009 \\
    Renting house  &  0.843 +/- 0.02 & 0.871 +/- 0.018 \\
    Selling apartment & 0.855 +/- 0.013  &  0.889 +/- 0.011 \\
    Selling house & 0.802 +/- 0.038  & 0.829 +/- 0.050  \\
    \hline
    \end{tabular}%
  \label{tab-res-svm-corr}%
\end{table}%

To understand why text mining reduced the RMSE, we investigated the linear regression model in Stage 2 to identify which words affected the price positively or negatively. Table \ref{tab-words} lists a sample of the discovered important words. Few of the words that affected the price are related to the location. While the location attribute in the original dataset\footnote{Recall that the location attribute was converted to binary attributes corresponding to each location value.} did specify the location of the unit in a structured manner, only one location was allowed. Textual description on the other hand allowed mentioning nearby prime locations and landmarks. For example, the \textbf{Palm} islands and \textbf{Downtown} are prime locations in Dubai, and therefore affect the price positively in several of the datasets. The \textbf{Burj} refers to Burj Khalifa, the tallest artificial structure in the world, which is a prime landmark with luxurious real estate surrounding it. The \textbf{Sports} city is a new real estate project that is not fully developed yet (hence the reduction in rent). Most of the words that affected the price negatively made sense, such as \textbf{deal, offer, road, price} and \textbf{sale}. Interestingly, some of the words affected the price negatively despite representing positive sentiment, such as \textbf{spacious} and \textbf{nice}. This is in contrast to words like \textbf{amazing, stunning} and \textbf{beautiful} that also represent positive sentiment but affected the price positively. A possible explanation is that only words with strong positive and confident sentiment affect the price positively, while words with weak positive sentiment reflect weaker confidence in the actual worth of the real estate property.  

\begin{table}[htbp]
  \centering
  \caption{Sample of words (uni-gram and bi-gram) that affect the price of a real estate classified (either positively or negatively)}
    \begin{tabular}{|r|c|c|}
    \hline
    Dataset & Positive words  & Negative Words \\
    \hline
    Renting  & study, suite  & sale, sports \\
Apartment & palm, amazing& nice, deal \\
    \hline
    Renting  & downtown, palm & partial, deal \\
House & luxurious, stunning & offer, village \\
    \hline
    Selling  & burj, quality& road, covered \\
Apartment & study, beautiful & cluster, plot\\
    \hline
    Selling  & proud, finishing &  hotel, fully \\
House	& golf, views& price , spacious\\
    \hline
    \end{tabular}%
  \label{tab-words}%
\end{table}%

\section{Related Work}
\label{sec-related}
Due to the importance of valuating a real estate property, there has been extensive research on automatic valuation
 \cite{rossini2000using,rossini1999accuracy,case2004modeling,bourassa2010predicting,stanley12jpr,Helbich201381}. Most of that work used hedonic models, which assumed the price can be predicted from a combination (usually linear) of the property structured features \cite{frew2003estimating} such as number of bedrooms and the location. Traditional hedonic models were based on human expertise, where the model parameters are usually hand-coded by experts, unlike our proposed method here, which uses data mining. There has been growing literature on the use of data mining techniques to analyze real estate data \cite{jaen02flairs,guan08jrer,chen2010application,gacovski2012data,helbich2013data}, however, most of the previous work focused only on structured features and ignored textual features. We review sample of these works in the remainder of this section. 

One of the early works \cite{jaen02flairs} used decision tree and neural network techniques  to predict the sale price of a house. The analysis used data with 15 numerical features that represent the houses’ characteristics plus a categorical feature that corresponds to the address. The dataset consisted of 1000 records that were collected from the houses’ sales transactions in Miami, US. Unlike our work, the analysis focused only on properties for sale (did not include rentals), used only structured features (no text mining) and relied on a much smaller dataset (compared to our +50,000 records).
A broader analysis was conducted in \cite{wedyawati04iciri}, covering 295,787 transactions from four cities in the US. Again, only numerical features were used (although more extensive features were used, almost 200) and no textual features were used (also despite attempting to predict the price, no performance criterion was reported). 
A more recent work \cite{guan08jrer} proposed Adaptive Neuro Fuzzy Inference System (ANFIS) and tested the system over 360 records of past sales properties in Midwest, US. The dataset had 14 numerical features and again no textual feature was used. 

Another research paper \cite{gacovski2012data} focused on studying the prediction of prices of apartments in city in Macedonia. Among the three data mining techniques that were applied on a dataset of 1200 sales transactions, the logistic regression (very similar to linear regression) was found to be the superior in prediction accuracy over decision tree and neural network techniques. Like the other earlier mentioned papers, there was no use of textual data. Some attempted to add structure to unstructured and ungrammatical data. However, this requires domain knowledge to build a reference structure (model) which can be used to extract the corresponding features  \cite{michelson08jair}. Our proposed approach does not require deep domain knowledge (aside from simple data cleansing, the whole process is mostly automated). 

The most related work to ours was concurrently and independently developed for analyzing real estate classifieds in the United States \cite{stevens14msc}. However, that work only used textual features along with structured features in one stage and therefore does not offer the flexibility and the semantic meaning (for keywords) as our 2-stage model.

\section{Conclusion and Future Work}
We propose in this paper a 2-stage regression model that uses text mining  to improve the prediction of the price of real estate classifieds. We show that using text mining significantly reduces the RMSE of prediction. We also show how our proposed approach can identify keywords that affect the price positively or negatively. 

One of the direction we want to pursue is extending our analysis to the Arabic language (which is commonly used in our region). We are also considering the integration of our system with a named-entity recognition component \cite{abdallah12cicling}, particularly for identify locations in ungrammatical text, to improve accuracy.

\bibliographystyle{abbrv}
\bibliography{references}

\begin{thebibliography}{10}

\bibitem{abdallah12cicling}
S.~Abdallah, K.~F. Shaalan, and M.~Shoaib.
\newblock Integrating rule-based system with classification for arabic named
  entity recognition.
\newblock In {\em Computational Linguistics and Intelligent Text Processing -
  13th International Conference, CICLing 2012, New Delhi, India, March 11-17,
  2012, Proceedings, Part {I}}, pages 311--322, 2012.

\bibitem{bourassa2010predicting}
S.~C. Bourassa, E.~Cantoni, and M.~Hoesli.
\newblock Predicting house prices with spatial dependence: A comparison of
  alternative methods.
\newblock {\em Journal of Real Estate Research}, 32(2):139--159, 2010.

\bibitem{case2004modeling}
B.~Case, J.~Clapp, R.~Dubin, and M.~Rodriguez.
\newblock Modeling spatial and temporal house price patterns: A comparison of
  four models.
\newblock {\em The Journal of Real Estate Finance and Economics},
  29(2):167--191, 2004.

\bibitem{chen2010application}
T.-H. Chen and C.-W. Chen.
\newblock Application of data mining to the spatial heterogeneity of foreclosed
  mortgages.
\newblock {\em Expert Systems with Applications}, 37(2):993--997, 2010.

\bibitem{crowston99ijem}
K.~Crowston and R.~T. Wigand.
\newblock Real estate war in cyberspace: An emerging electronic market?
\newblock {\em International Journal of Electronic Markets}, 9(1-2):1--8, 1999.

\bibitem{frew2003estimating}
J.~Frew and G.~Jud.
\newblock Estimating the value of apartment buildings.
\newblock {\em Journal of Real Estate Research}, 25(1):77--86, 2003.

\bibitem{gacovski2012data}
Z.~Gacovski, J.~Kolic, R.~Dukova, and M.~Markovski.
\newblock Data mining application for real estate valuation in the city of
  skopje.
\newblock {\em ICT Innovations 2012, Web Proceedings ISSN 1857-7288}, pages
  537--538, 2012.

\bibitem{guan08jrer}
J.~Guan, J.~Zurada, and A.~S. Levitan.
\newblock {An Adaptive Neuro-Fuzzy Inference System Based Approach to Real
  Estate Property Assessment}.
\newblock {\em Journal of Real Estate Research}, 30(4):395--422, 2008.

\bibitem{hansen1990neural}
L.~K. Hansen and P.~Salamon.
\newblock Neural network ensembles.
\newblock {\em IEEE Transactions on Pattern Analysis \& Machine Intelligence},
  (10):993--1001, 1990.

\bibitem{helbich2013data}
M.~Helbich, W.~Brunauer, J.~Hagenauer, and M.~Leitner.
\newblock Data-driven regionalization of housing markets.
\newblock {\em Annals of the Association of American Geographers},
  103(4):871--889, 2013.

\bibitem{Helbich201381}
M.~Helbich, A.~Jochem, W.~Mücke, and B.~Höfle.
\newblock Boosting the predictive accuracy of urban hedonic house price models
  through airborne laser scanning.
\newblock {\em Computers, Environment and Urban Systems}, 39(0):81 -- 92, 2013.

\bibitem{humphries2012automatically}
S.~Humphries, D.~Xiang, J.~Burstein, Y.~Bun, and J.~Ultis.
\newblock Automatically determining a current value for a home, Mar.~20 2012.
\newblock US Patent 8,140,421.

\bibitem{jaen02flairs}
R.~D. Jaen.
\newblock Data mining: An empirical application in real estate valuation.
\newblock In S.~M. Haller and G.~Simmons, editors, {\em FLAIRS Conference},
  pages 314--317. AAAI Press, 2002.

\bibitem{stanley12jpr}
S.~McGreal and P.~Taltavull~de La~Paz.
\newblock An analysis of factors influencing accuracy in the valuation of
  residential properties in spain.
\newblock {\em Journal of Property Research}, 29(1):1--24, 2012.

\bibitem{michelson08jair}
M.~Michelson and C.~A. Knoblock.
\newblock Creating relational data from unstructured and ungrammatical data
  sources.
\newblock {\em Journal of Artificial Intelligence Research(JAIR)}, 31:543--590,
  2008.

\bibitem{pera13is}
M.~S. Pera, R.~Qumsiyeh, and Y.-K. Ng.
\newblock Web-based closed-domain data extraction on online advertisements.
\newblock {\em Information Systems}, 38(2):183--197, Apr. 2013.

\bibitem{rossini1999accuracy}
P.~Rossini.
\newblock Accuracy issues for automated and artificial intelligent residential
  valuation systems.
\newblock In {\em International Real Estate Society Conference}, 1999.

\bibitem{rossini2000using}
P.~Rossini et~al.
\newblock Using expert systems and artificial intelligence for real estate
  forecasting.
\newblock In {\em Sixth Annual Pacific-Rim Real Estate Society Conference,
  Sydney, Australia}, pages 24--27. Citeseer, 2000.

\bibitem{shevade2000improvements}
S.~K. Shevade, S.~S. Keerthi, C.~Bhattacharyya, and K.~R.~K. Murthy.
\newblock Improvements to the smo algorithm for svm regression.
\newblock {\em Neural Networks, IEEE Transactions on}, 11(5):1188--1193, 2000.

\bibitem{stevens14msc}
D.~Stevens.
\newblock {Predicting Real Estate Price Using Text Mining}.
\newblock Master's thesis, Tilburg University School of Humanities, the
  Netherlands, 2014.

\bibitem{wedyawati04iciri}
W.~Wedyawati and M.~Lu.
\newblock Mining real estate listings using {ORACLE} data warehousing and
  predictive regression.
\newblock In {\em Proceedings of the 2004 {IEEE} International Conference on
  Information Reuse and Integration, {IRI} - 2004, November 8-10, 2004, Las
  Vegas Hilton, Las Vegas, NV, {USA}}, pages 296--301, 2004.

\end{thebibliography}

\end{document}